\documentclass[onecolumn,aps,superscriptaddress,nofootinbib]{revtex4-2}
\usepackage{newtxtext}
\usepackage[varvw,bigdelims]{newtxmath}
\usepackage[colorlinks=true,allcolors=blue]{hyperref}
\usepackage{graphicx}
\usepackage{bm}
\usepackage{amsmath}
\usepackage{xcolor}
\usepackage{array}
\usepackage{multirow}

\begin{document}
\title{Feasibility of the observation of $\eta^{\prime}$ mesic nuclei in the semi-exclusive $^{12}$C($p, dp$) reaction}

\author{Natsumi Ikeno}
\email{ikeno@maritime.kobe-u.ac.jp}
\affiliation{Department of Agricultural, Life and Environmental Sciences, Tottori University, Tottori 680-8551, Japan}
\affiliation{Graduate School of Maritime Sciences, Kobe University, Kobe 658-0022, Japan}

\author{Yuko Higashi}
\affiliation{Department of Physics, Nara Women's University, Nara 630-8506, Japan}

\author{Hiroyuki Fujioka}
\affiliation{Department of Physics, School of Science, Institute of Science Tokyo, Meguro, Tokyo 152-8551, Japan}

\author{Kenta Itahashi}
\affiliation{RIKEN Nishina Center for Accelerator-Based Science, RIKEN, 2-1 Hirosawa, Wako, Saitama 351-0198, Japan}
\affiliation{RIKEN Cluster for Pioneering Research, RIKEN, Wako, Saitama 351-0198, Japan}
\affiliation{Department of Physics, The University of Osaka, Toyonaka, 560–0043 Osaka, Japan}

\author{Ryohei Sekiya}
\affiliation{Department of Physics, Kyoto University, Kitashirakawa-Oiwake, Sakyo, Kyoto 606-8502, Japan}

\author{Yoshiki K. Tanaka}
\affiliation{RIKEN Cluster for Pioneering Research, RIKEN, Wako, Saitama 351-0198, Japan}

\author{Junko Yamagata-Sekihara}
\affiliation{Department of Physics, Kyoto Sangyo University, Kyoto 603-8555, Japan}

\author{Volker Metag} 
\affiliation{Universit\"{a}t Giessen, Heinrich-Buff-Ring 16, 35392 Giessen, Germany}

\author{Mariana Nanova}
\affiliation{Universit\"{a}t Giessen, Heinrich-Buff-Ring 16, 35392 Giessen, Germany}

\author{Satoru Hirenzaki}
\affiliation{Department of Physics, Nara Women's University, Nara 630-8506, Japan}

\date{\today}

\begin{abstract}%
We study theoretically the feasibility of the semi-exclusive $^{12}$C($p,dp$)$X$ reaction for the observation of $\eta^\prime$ mesic nuclei using the microscopic transport model JAM. The semi-exclusive measurements of the ($p,d$) reaction with protons from $\eta^\prime$ absorption are found to be significant for the observation of the $\eta^\prime$ bound states. 
Especially, the measurements of the energetic protons from $\eta^\prime$ non-mesic two-body absorption ($\eta^\prime NN \to NN$) are considered to be critically important.
The Green's function method is used to calculate the expected spectrum of forward going deuterons corresponding to the excitation energy spectrum of the $\eta^\prime \otimes {}^{11}$C system in the semi-exclusive measurement.
The semi-exclusive measurements are shown to be important in general for the $\eta^\prime$ mesic nucleus observation.
\end{abstract}

\maketitle
\section{Introduction}
Studies of the meson-nucleus systems and the in-medium properties of hadrons are one of the interesting subjects in contemporary hadron-nuclear physics to investigate the nature of QCD in the low energy region and the aspects of symmetry at finite nuclear density $\rho$ \cite{Batty:1997zp,Friedman:2007zza,Hayano:2008vn,Metag:2017yuh}. For example, spectroscopic studies of the pionic atoms provide important information on the change of the chiral condensate and the partial restoration of chiral symmetry at finite $\rho$~\cite{Yamazaki:2012zza,Hirenzaki:2022dpt,Itahashi:2023boi,piAF:2022gvw}.

We consider in this article the formation of an $\eta^\prime(958)$ ($\eta^\prime$) meson-nucleus bound state ($\eta^\prime$ mesic nucleus)~\cite{Nagahiro:2004qz,Nagahiro:2006dr,Cobos-Martinez:2023hbp,Nagahiro:2012aq,Itahashi:2012ut,Jido:2011pq}. Theoretically, the origin of the exceptionally large $\eta^\prime$ mass is attributed to the interplay of chiral symmetry breaking and $U_A$(1) anomaly~\cite{Jido:2011pq,Nagahiro:2012aq,Kunihiro:1989my,Hatsuda:1994pi,Bass:2018xmz}. Thus, the study of the $\eta^\prime$ mesic nucleus will provide unique information on the effects of the $U_A(1)$ anomaly at finite density, which would be difficult to obtain from studies of other light pseudo-scalar octet mesons.
The modifications of the in-vacuum $\eta^\prime$--nucleon interaction in nuclei~\cite{Sakai:2022xao,Ikeno:2025kwe} and the $\eta^\prime$--nucleus optical potential~\cite{Nagahiro:2011fi,Friedman:2025avs} are also quite interesting.
The formation of $\eta^\prime$ bound states were studied theoretically in Refs.~\cite{Nagahiro:2004qz,Nagahiro:2006dr,Nagahiro:2012aq,Itahashi:2012ut,Miyatani:2016xfq}, and the results of the formation experiments by the ($p,d$) reaction were reported in Refs.~\cite{n-PRiMESuper-FRS:2016vbn,e-PRiMESuper-FRS:2017bzq,Sekiya:2025hwz}.
Some indications of $\eta^\prime$ bound states were found in Ref.~\cite{Sekiya:2025hwz}, although the statistical significance was not sufficient.
The results of the ($\gamma,p$) reaction, looking for back-to-back $\eta$--$p$ pairs from the decay of $\eta^\prime$ mesic states, were published in Ref.~\cite{LEPS2BGOegg:2020cth}.
Results on the photo-production of $\eta^\prime$ mesons off nuclei and the interaction between $\eta^\prime$ mesons and nuclei at higher energies were reported in Refs.~\cite{Friedrich:2016cms,CBELSATAPS:2016qdi}.

In this article, we discuss the semi-exclusive ${}^{12}$C($p,dp$) reaction theoretically to reduce the contributions from the background processes and to observe the signals of the $\eta^\prime$ mesic nucleus formation more clearly~\cite{Fujioka:2014nfa,Super-FRS:2015sjd,Higashi:Mthesis}.
In Section~\ref{sec:2}, we explain the $^{12}\mathrm{C}(p,d)$ spectrum for the formation of the $\eta^\prime$ bound states using the theoretical spectrum obtained by the Green's function method.  
In Section~\ref{sec:3}, we show the results of the semi-exclusive $^{12}\mathrm{C}(p,d\,p)$ spectra obtained by the JAM simulation, and explain the advantages of the semi-exclusive observation.  
Section~\ref{sec:4} is devoted to the conclusions.  
In Appendix~\ref{sec:A}, the relative probabilities of the energetic hadron productions are estimated for each $\eta^\prime$ absorption process in the nucleus using simple models.  
In Appendix~\ref{sec:B}, the variables appeared in the simulations in this article are compiled in a Table.

\section{Theoretical $^{12}$C($p,d$) spectrum by Green's function method }\label{sec:2}
We show firstly the theoretical $^{12}$C($p,d$) spectrum for the formation of the $\eta^\prime$ mesic nucleus calculated by the Green's function method~\cite{Nagahiro:2012aq,Yamagata:2007cp}.
We consider a phenomenological $\eta^\prime$--nucleus optical potential $U_{\eta^{\prime}}$ as
\begin{equation}
 U_{\eta^{\prime}}(r) = V \left( \frac{\rho(r)}{\rho_0} \right)^1
+ i \ W_1 \left( \frac{\rho(r)}{\rho_0} \right)^1
+ i \ W_2 \left( \frac{\rho(r)}{\rho_0} \right)^2,
\label{eq:U_etap}
\end{equation}
where $\rho(r)$ indicates the density distribution of the daughter nucleus $^{11}$C and $\rho_0$ the normal nuclear density $\rho_0 = 0.17$~fm$^{-3}$. 
The density $\rho(r)$ is fixed to be 
$\displaystyle \rho(r) = \rho(0) (1+ a({r}/{c})^2) \exp(-(r/c)^2)$ 
with the parameters $a=0.811$ and $c = 1.690$~fm assuming the same form as that of $^{11}$B in Ref.~\cite{DeJager:1974liz}.
The density of $^{12}$C is taken to be the same functional form as $^{11}$C with the parameters $a=1.150$ and $c = 1.672$~fm~\cite{DeJager:1974liz}.
The real part of the potential is assumed to be attractive and provides the $\eta^\prime$ bound states. 
The strength of the real potential at $\rho_0$ is determined by the parameter $V$.
The imaginary parts of the potential describe the absorptive processes of $\eta^\prime$ in the nucleus, and in other words, the decay processes of the $\eta^\prime$ mesic nucleus.
The parameter $W_1$ determines the strength of the $\eta^\prime$ absorption by one-body processes such as $\eta^\prime N \to \pi N$ and $\eta^\prime N \to \eta N$, and the parameter $W_2$ determines the $\eta^\prime$ absorption strength by two-body processes including the non-mesic decay process $\eta^\prime NN \to NN$.

\begin{figure}[!htb]
\begin{center}
\includegraphics[scale=0.4]{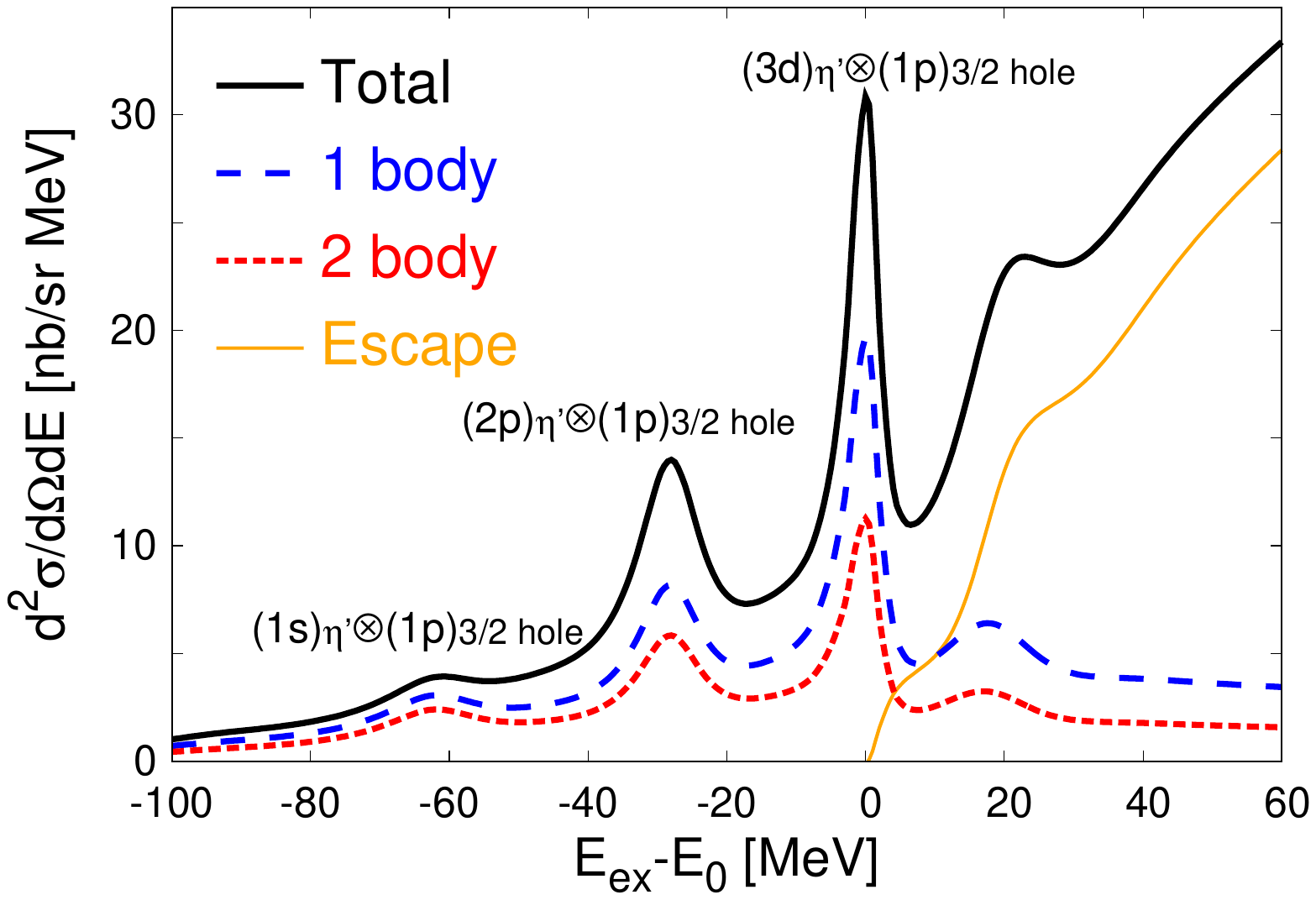}
\caption{
Calculated spectra of the $^{12}$C($p,d$)$^{11}$C$\otimes~\eta^\prime$ reaction for the formation of $\eta^\prime$--nucleus systems with proton kinetic energy $T_p = 2.5$~GeV and deuteron angle $\theta_d =0^\circ$ as a function of the excitation energy $E_\text{ex}$. $E_0$ is the $\eta^\prime$ production threshold. The $\eta^\prime$--nucleus optical potential is given in Eq.~\eqref{eq:U_etap}. The potential parameters are taken to be $(V, W_1, W_2) = (-100, -5 ,-5)$~MeV.
The thick solid line shows the total spectrum. The dashed and dotted lines show the contributions from the one-body and two-body $\eta^\prime$ absorption processes, respectively, calculated as the conversion parts. The contribution from the $\eta^\prime$ escape process is also shown.
 }
 \label{fig:1}
\end{center}
\end{figure}

In the Green's function method, the contributions of the absorptive processes to the excitation energy spectrum can be calculated as the conversion parts. The contribution from the $\eta^\prime$ escape process can also be evaluated separately. We show the calculated spectrum in Fig.~\ref{fig:1}, for a set of the potential parameters $(V, W_1, W_2) = (-100, -5 ,-5)$~MeV. 
As research on the optical potential has been insufficient to date, and uncertainties remain, we can only make assumptions about the shape and strength of the potential here. Please note that the ratio of $W_1$ and $W_2$ of the assumed potential is comparable to a theoretical potential~\cite{Nagahiro:2011fi} with an absolute value of the scattering length of $0.5~\mathrm{fm}$. The strengths of the real and the imaginary parts of the potential are also comparable to the theoretical potential~\cite{Nagahiro:2011fi} with different values of model parameters.

As we can see from Fig.~\ref{fig:1}, each conversion part maintains almost the same spectral shape and peak structures as the total spectrum in the energy range of bound state formation, so that the information on the bound states can be obtained by observing the conversion spectrum. 
These conversion parts of the deuteron spectrum are obtained experimentally as the semi-exclusive cross sections, where the additional protons emitted by the one-body and two-body $\eta^\prime$ absorption processes are measured in coincidence. 
Thus, in case that the semi-exclusive measurements of the conversion spectra effectively suppress the background and thereby significantly improve  the signal-to-background ($S/B$) ratio, the semi-exclusive measurements of the conversion spectra are quite useful to find experimental signals of the $\eta^\prime$ mesic nucleus formation.

\section{Simulation of the semi-exclusive $^{12} \text{C}(p, dp)X$ reaction}\label{sec:3}
We consider the semi-exclusive $^{12} \text{C}(p, dp)$ reaction, where the additional protons emitted by the $\eta^\prime$ meson absorption are measured in coincidence, and investigate whether the semi-exclusive measurements are really effective or not to reduce the background for the clear observation of the $\eta^\prime$ mesic nucleus formation.
We think that the background contributions to the excitation energy spectrum can be significantly suppressed by the semi-exclusive measurements using the protons from the $\eta^\prime$ absorption in the backward directions, as explained below.
In the background processes, dominated by multi-pion production, the particles are emitted mainly to forward angles because of the energetic incident proton beam, while the emitted particles from the absorption of the bound $\eta^\prime$ are expected to have a uniform angular distribution in the $\eta^\prime$ mesic nucleus rest system.
Since the recoil momentum $q$ of the mesic nucleus formed in the ${}^{12}\mathrm{C}(p,d){}^{11}\mathrm{C} \otimes \eta^\prime$ reaction at an incident proton energy of 2.5 GeV is as small as $q \sim 0.4~\mathrm{GeV}/c$~\cite{Nagahiro:2012aq}, corresponding to a recoil velocity of only 0.3\% of the speed of light, it is negligible in the present study.  Therefore, the particles emitted in the $\eta^\prime$ absorption process can be considered to have a nearly uniform angular distribution in the laboratory system.
Thus, the observation of the additional particles at backward angles is considered to be most efficient to reduce the background contributions to the excitation energy spectrum.
The momenta of the emitted protons $p_p$ after $\eta^\prime$ absorption by a nucleon at rest are considered to be 
$p_p \simeq 0.58$~GeV/$c$ for $\eta^\prime N \to \eta N$ and
$p_p \simeq 0.71$~GeV/$c$ for $\eta^\prime N \to \pi N$ in one-body absorption processes as indicated in Refs.~\cite{Fujioka:2014nfa,Super-FRS:2015sjd}.
The protons emitted in $\eta^\prime $ absorption by the non-mesic two-body process $\eta^\prime NN \to NN$ by two rest nucleons are expected to have 
a large kinetic energy 
$\displaystyle \sim \frac{m_{\eta^\prime}}{2} $, 
and, thus,
a larger momentum of 
$p_p \simeq 1.0~$GeV/$c$.
We expect that the protons with larger momenta emitted in the two-body absorption process at backward angles could be distinguished most clearly from those from the background processes.
Thus, we begin by noting the $\eta^\prime NN \to NN$ process as an important candidate for the source of protons from the signal of the semi-exclusive measurements and perform the numerical simulation in Sect.~\ref{sect:3.1}. Then, in Sect.~\ref{sect:3.2} we consider both one- and two-body absorption processes of $\eta^\prime$ as the decay of the $\eta^\prime$ mesic nucleus.
The idea of this semi-exclusive measurements was proposed in Refs.~\cite{Fujioka:2014nfa,Super-FRS:2015sjd} and the preliminary results of the simulation were reported in Ref.~\cite{Higashi:Mthesis}.
To evaluate the advantage of the semi-exclusive reaction quantitatively, we perform the simulation using the microscopic transport model JAM~\cite{Nara:1999dz}.

\begin{figure*}[tb!]
 \begin{center}
\includegraphics[scale=0.52]{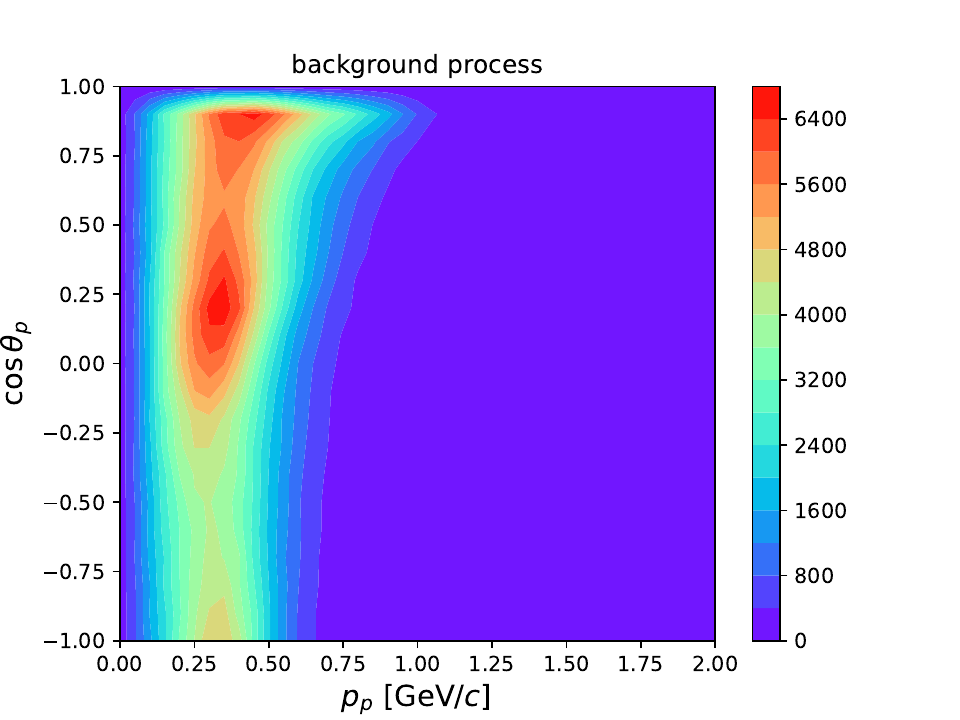} 
\includegraphics[scale=0.52]{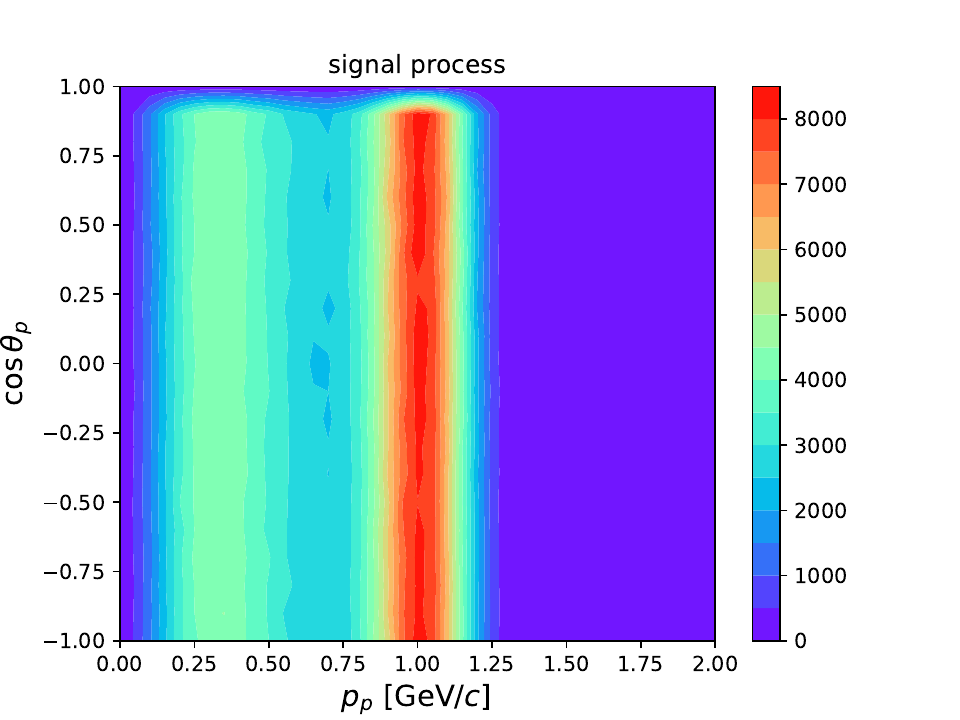}
  \caption{
The distributions of protons emitted in the inclusive $^{12}$C($p,d$)$X$ reaction in the plane of the proton momentum $p_p$ and $\cos \theta_p$ of the proton emission angle $\theta_p$ in the laboratory system in two independent simulations
for the incident proton kinetic energy $T_p = 2.5$~GeV.
The simulated result of the background processes is shown in (left), and that of the signal process from the $\eta^\prime$ mesic nucleus decay by the two-body absorption $\eta^\prime NN \to NN$ is shown in (right).
}
  \label{fig:3}
 \end{center}
\end{figure*}

As background processes, we calculate in this article the $^{12}$C($p,d$)$X$ inclusive reaction with a forward going deuteron. The simulation is performed for $1.44 \times 10^{11}$ incident protons.
The results of the excitation energy momentum spectrum are consistent with the experimental data in Refs.~\cite{n-PRiMESuper-FRS:2016vbn,e-PRiMESuper-FRS:2017bzq}.
This simulation does not include the contributions of the $\eta^\prime$ mesic nucleus formation and corresponds to background processes. 
We show in Fig.~\ref{fig:3} (left) the distribution of emitted protons  in this JAM simulation for events in the excitation energy range $-60$~MeV$\le E_\text{ex} - E_0 \le 60$~MeV in the plane of the proton momentum $p_p$ and $\cos \theta_p$ of the proton emission angle $\theta_p$. $E_0$ is the $\eta^\prime$ production threshold. As shown in this figure, most protons from the background in the $^{12}$C($p,d$)$X$ reaction tend to be distributed in forward direction and do not have large momenta $p \gtrsim 0.5$~GeV/$c$ at backward angles as we expect.

In the following subsections, we consider the proton distribution from the signal processes, which are defined as the formation of the $\eta^\prime$ mesic nucleus followed by the decay by the $\eta^\prime$ absorption processes.
The deuteron spectra of the signal processes are corresponding to the conversion parts in Fig.~\ref{fig:1}.
We perform the intra-nuclear cascade simulation independent from that of the background using JAM by placing energetic particles such as nucleon(s) and $\pi$/$\eta$ inside the nucleus which are produced by the $\eta^\prime$ absorption processes. These particles are prepared in the simulation randomly with the spatial distribution proportional to $\rho(r)$ and $\rho^2(r)$ of $^{11}$C for the one- and two-body absorption, respectively, and with the momentum distributions centered on the values corresponding to the $\eta^\prime$ absorptions by the nucleon(s) at rest.
The effects of the Fermi motion of nucleon(s) and those of the bound $\eta^\prime$ momentum distribution before absorption are taken into account in calculating the momentum distribution of energetic particles produced by absorption. The number of forward going deuterons is the same as that of $\eta^\prime$ mesic nuclei populated in the signal process.

In the appendix~\ref{sec:A}, the relative probabilities of the energetic proton and/or neutron production are evaluated for each absorption process of the bound $\eta^\prime$ by nucleon(s) assuming the simple absorption mechanisms.
The relative probabilities of the energetic $\pi$ and $\eta$ production are also evaluated for one-body absorption of $\eta^\prime$.

\subsection{Protons from $\eta^\prime$ mesic nuclei by non-mesic two-body absorption of $\eta^\prime$ }\label{sect:3.1}

We consider the $\eta^\prime NN \to NN$ non-mesic two-body absorption of $\eta^\prime$ for the decay of the $\eta^\prime$ mesic nucleus here.
We perform the intra-nuclear cascade simulation by assuming the existence of the $\eta^\prime$ mesic nuclei. 
The number $N_0$ of forward deuterons is the same as the number of the $\eta^\prime$ mesic nuclei in the signal process.
The nucleons are supposed to have the momentum $p_p = 1.04$~GeV/$c$ as the central value inside nucleus after $\eta^\prime$ absorption. 
Two nucleons after $\eta^\prime$ absorption by $\eta^\prime NN \to NN$ have momenta of the same magnitude with the opposite direction in their center of mass system.
After the non-mesic two-body absorption, three charged states of two nucleon, $pp$, $pn$, and $nn$, are possible. In appendix~\ref{sec:A}, we show an estimation of the probabilities $P_{pp}$, $P_{pn}$, and $P_{nn}$ of appearance of these final two nucleon states assuming the simple absorption mechanisms. Thus, for $N_0$ $\eta^\prime$ mesic nuclei, we have energetic $N_0 (2P_{pp} + P_{pn})$ protons and $N_0 (2 P_{nn} + P_{pn})$ neutrons with the momentum around 1~GeV/$c$ inside the nucleus after two-nucleon absorption.
We consider all these energetic protons and neutrons, and simulate the distribution of protons emitted from the nucleus. The energetic neutrons can be the source for proton emission. Fig.~\ref{fig:3} (right) shows the simulated distribution of protons emitted in the $\eta^\prime$ absorption by the two-body process $\eta^\prime NN \to NN$ as the indication for the formation of the $\eta^\prime$ mesic nucleus.
We find that the proton distribution is very different from that of the background processes and has a peak at $p_p \sim 1$~GeV/$c$ for all emitted angles as expected.

The clear difference of the simulation results for the proton distributions from the signal and background processes enables us to find out suitable condition of the semi-exclusive ($p, dp$) reaction for observing a clearer signal for the formation of the $\eta^\prime$ mesic state in $^{11}$C. From Fig.~\ref{fig:3}, it is clear that protons with higher momenta emitted to backward angles should be detected selectively in the semi-exclusive measurements to improve the $S/B$ ratio of the excitation energy spectrum.
To show the advantage of the semi-exclusive measurements, we define the improvement factor $f_{S/B}$ of the $S/B$ ratio of the excitation energy spectra as the ratio of the number of coincidences requesting the forward deuterons and proton from the signal ($N_{\text{c}}^S$) and from the background ($N_{\text{c}}^B$) processes, normalized to 1 for the inclusive ($p,d$) reaction obtained by the forward deuteron numbers from the signal ($N_d^S$) and the background ($N_d^B$) processes as,
\begin{equation}
f_{S/B} = \left. \left(\frac{N_{\text{c}}^S}{N_{\text{c}}^B}\right) \middle/  \left( \frac{N_d^S}{N_d^B}\right)_\text{inclusive} \right. \;.
\label{eq:SB}
\end{equation}
By imposing cut conditions on the momenta and angles of the emitted protons, both $N_{\text{c}}^S$ and $N_{\text{c}}^{B}$ decrease, but the differences in the proton distributions of the signal and the background can magnify the $f_{S/B}$ factor for suitable cut conditions. At the same time, the cut conditions reduce the number of events and make the statistics worse. Thus, the best experimental cut condition will be determined by optimizing these effects.

For the simulation of the background, 
the number of the coincidence $N_{\text{c}}^B$ for semi-exclusive reactions is defined as the number of forward deuterons with at least one proton emission which satisfies the cut conditions. 
For the simulation of the signal processes, we take care of the possibility to have multi-proton emissions from one $\eta^\prime$ mesic nucleus, namely from one nucleon pair, corresponding to one deuteron forward. 
Using the probabilities $P_{pp}$, $P_{pn}$, and $P_{nn}$ of producing energetic $pp$, $pn$, and $nn$ pairs inside the nucleus, we determine the number of three kinds of nucleon pairs after $\eta^\prime$ absorption. 
Then, we perform the JAM simulation to obtain the number of the nucleon pairs of the production of only one emitted proton outside the nucleus which satisfies the cut conditions.
Similarly, we also evaluate by JAM simulations the number of pairs of emitted protons satisfying the cut conditions. The latter events are used to evaluate the strength of the back-to-back correlation of the emitted protons from the $\eta^\prime$ mesic nucleus decay.
We evaluate the number of forward deuterons accompanying at least one proton satisfying the experimental cut conditions 
as the sum of the number of the nucleon pairs with only one emitted proton and those with two or more emitted protons satisfying the cut conditions.

We show in Table~\ref{tab:1} the results of the $f_{S/B}$ factor for the simulations of various cut conditions of the proton momentum $p_p$ and the proton emission angle $\theta_p$.
We first define the lowest value of $p_p$ and the smallest (largest) value of $\theta_p$ ($\cos\theta_p$) as the cut conditions. 
We find that the $S/B$ ratio is improved much in the semi-exclusive reaction, and that the $f_{S/B}$ factor is significantly larger than 1 for the stronger cut conditions. For example, if we consider protons emitted with momenta $p_p$ larger than 0.8~GeV/$c$ and at angles $\theta_p$ larger than 90$^\circ$ in the semi-exclusive measurements, the $S/B$ ratio is found to be about 200 times larger than that of the inclusive spectra.
Consequently, if the signal cross section were of the order of 1000~times smaller than the inclusive cross section, the $S/B$ ratio would still be of the order of 20\%. 

As shown in Table~\ref{tab:1}, the $f_{S/B}$ factor is found to be sensitive to the cut condition for $p_p$. 
Hence, we also show the results for 200~MeV/$c$ wide momentum bins and for backward angles. 
We find that the $f_{S/B}$ factor is almost 1 for the region with the lower value momentum $0.3 < p_p < 0.5$~GeV/$c$ because of the large background contributions. 
For the high momentum region  $0.9 < p_p < 1.1$~GeV/$c$,
 the $f_{S/B}$ factor is large as $4.2 \times 10^2$
 due to the large contributions from the two-body absorption of $\eta^\prime$.

\begin{table}[!tb]
\caption{\label{tab:1} 
Improvement factor $f_{S/B}$ of the $S/B$ ratio defined in Eq.~\eqref{eq:SB} for non-mesic two-body absorption of $\eta^\prime$ by the 
semi-exclusive $^{12}$C($p,dp$) reaction with respect to the inclusive  $^{12}$C($p,d$) reaction for the incident proton kinetic energy $T_p = 2.5$~GeV.
The improvement factor is normalized to 1 for the inclusive reaction
 where no cut condition for emitted protons is imposed.
In the semi-exclusive $^{12}$C($p,dp$) reaction considered here, protons in the final state are observed at backward angles with high momenta. The lowest value of the proton momentum $p_p$ and the largest value of $\cos \theta_p$ of the proton emission angle $\theta_p$ are shown as the cut conditions. 
Events with protons with higher momenta and larger emission angles (smaller $\cos \theta_p$ values) than the cut conditions are accumulated for both background and signal processes and used to evaluate the improvements of the $S/B$ ratio.
The results for the definite value of the proton momentum $p_p$ with the range of 200~MeV/$c$ are also shown for the backward angles $0.0 > \cos \theta_p$.
 }
\begin{tabular}{l||ccccc}
\hline
 &   \multicolumn{5}{c}{Proton momentum cut  [GeV/$c$]} \\ 
Proton angle cut &~~~~~~$0.6 < $~~~~& ~~~~$0.7 <$~~~~~~ & ~~~~$0.8 <$~~~~~~  & ~~~~$0.9< $~~~~~~  & ~~~~$1.0 < p_p$~~~~  \\
\hline
$\ ~0.50>  \cos \theta_p$ & $1.7 \times 10^1$ & $4.1 \times 10^1$ & $1.1 \times 10^2$ & $2.8 \times 10^2$ & $7.0 \times 10^2$  \\
$\ ~0.25>$ & $2.2 \times 10^1$ & $5.8 \times 10^1$ & $1.6 \times 10^2$ & $4.2 \times 10^2$ & $1.2 \times 10^3$  \\
$\ ~0.00>$ & $2.7 \times 10^1$ & $7.4 \times 10^1$ & $2.1 \times 10^2$ & $5.8 \times 10^2$ & $1.8 \times 10^3$  \\
$-0.25>$ & $3.1 \times 10^1$ & $8.6 \times 10^1$ & $2.5 \times 10^2$ & $6.9 \times 10^2$ & $2.2 \times 10^3$  \\
$-0.50>$ & $3.3 \times 10^1$ & $9.4 \times 10^1$ & $2.7 \times 10^2$ & $8.0 \times 10^2$ & $2.7 \times 10^3$  \\
$-0.75>$ & $3.5 \times 10^1$ & $1.0 \times 10^2$ & $3.0 \times 10^2$ & $9.8 \times 10^2$ & $4.1 \times 10^3$  \\
\hline
\hline\\
  & $0.3 < p_p < 0.5$& $0.5 < p_p < 0.7$ & $0.7 < p_p < 0.9$ & $0.9< p_p < 1.1$   \\
\hline
$\ ~0.0>  \cos \theta_p$~~~~~ & $1.1 \times 10^{0}$ & $2.3 \times 10^0$ & $1.9 \times 10^1$ & $4.2 \times 10^2$   \\
\hline
\end{tabular}
\end{table}

\subsection{Protons from $\eta^\prime$ mesic nuclei by the one- and two-body absorption of $\eta^\prime$}\label{sect:3.2}
\begin{figure*}[tb!]
 \begin{center}
\includegraphics[scale=0.5]{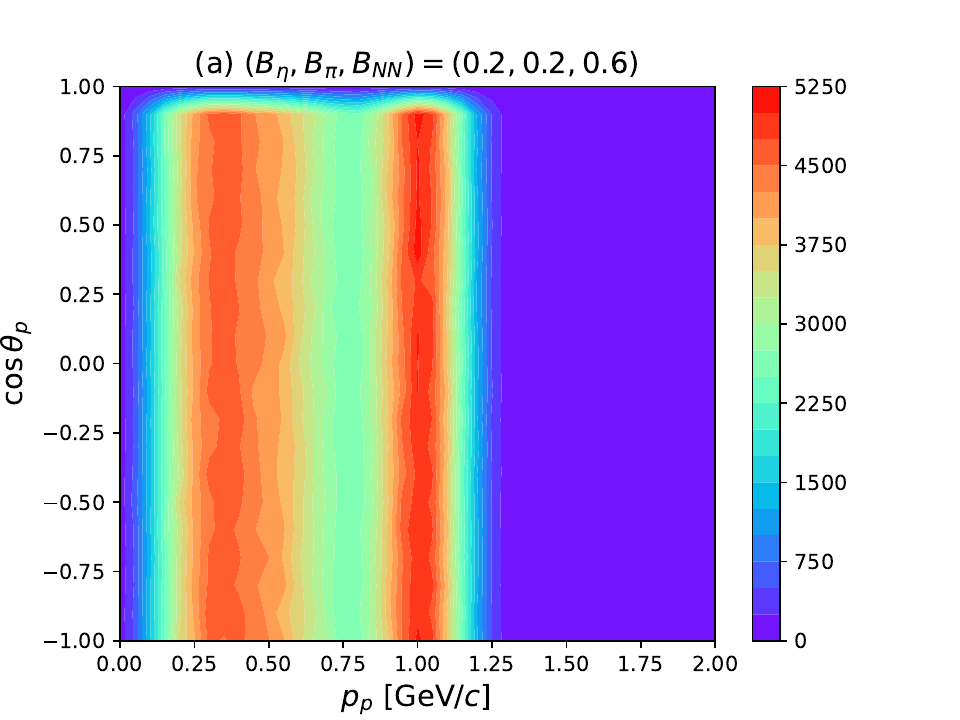} \
\includegraphics[scale=0.5]{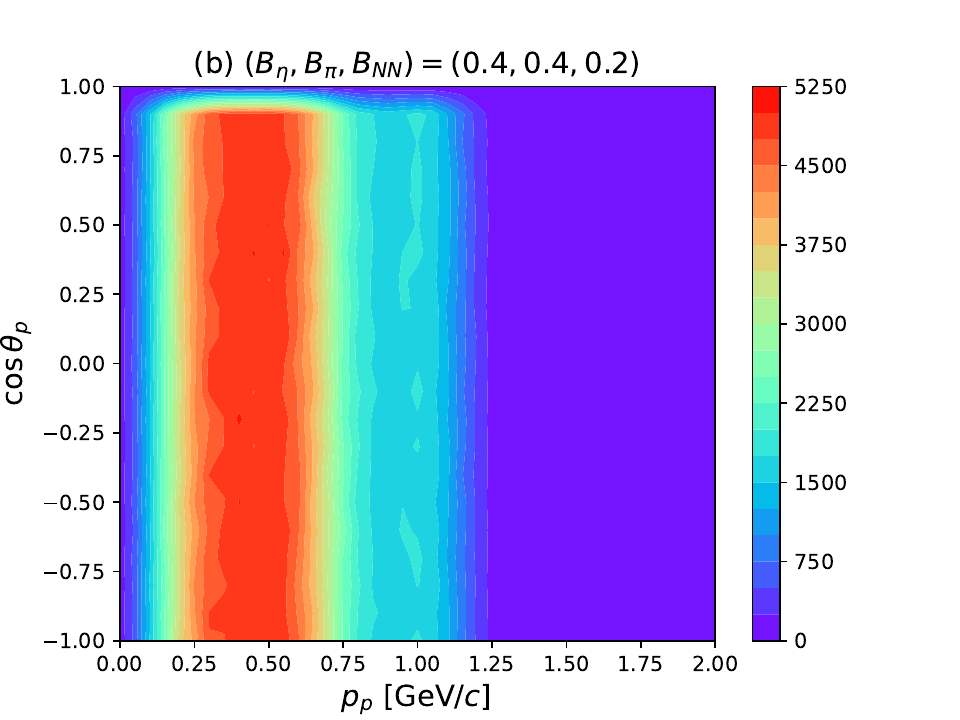} 
  \caption{
The distributions of protons emitted from the decays of the $\eta^\prime$ mesic nuclei, which are supposed to be populated in the $^{12}\text{C}(p,d)X$ reaction, in the plane of the proton momentum $p_p$ and $\cos \theta_p$ of the proton emission angle $\theta_p$ in the laboratory system. The decay branching ratios $B_\eta$, $B_\pi$, and $B_{NN}$ for $\eta^\prime N \to \eta N$, $\eta^\prime N \to \pi N$ and $\eta^\prime NN \to NN$ processes are assumed to be
(a) $(B_{\eta},  B_{\pi},  B_{NN}) = (0.2, 0.2, 0.6)$,  and
(b) $= (0.4,0.4, 0.2)$, respectively. 
}
  \label{fig:SBR}
 \end{center}
\end{figure*}

In this subsection, we perform  the more realistic simulation of the  $^{12}$C($p,dp$) reaction than sect.~\ref{sect:3.1}. We consider $\eta^\prime N \to \eta N$ and $\eta^\prime N \to \pi N$ processes as the one-body $\eta^\prime$ absorption in addition to $\eta^\prime NN \to NN$  two-body $\eta^\prime$ absorption discussed in the previous subsection.
The expected momentum distributions of the produced energetic protons after one-body $\eta^\prime$ absorption processes in nucleus have smaller central values than those of non-mesic two-body process as mentioned in the beginning of Sect.~\ref{sec:3}.
We consider all combinations of energetic particles produced by $\eta^\prime$ one-body absorption as shown in Appendix~\ref{sec:A}, namely, $\eta p$, $\eta n$, $\pi^- p$, $\pi^0 p$, $\pi^+ n$, and $\pi^0 n$.
In order to perform the intra-nuclear cascade simulation of all decay processes of the $\eta^\prime$ mesic nucleus considered in the article, we assume the decay branching ratios of the $\eta^\prime$ mesic nuclei $B_{\eta}$, $B_{\pi}$, and $B_{NN}$ for $\eta^\prime N \to \eta N$, $\eta^\prime N \to \pi N$ and $\eta^\prime NN \to NN$ processes, 
and combine the results of the simulations to obtain the results of the emitted proton distribution.
The numbers of $\eta^\prime$ mesic nucleus decaying into each process are $N_0 B_\eta$, $N_0 B_\pi$, and $N_0 B_{NN}$ for $N_0$ $\eta^\prime$ mesic nuclei with the condition $B_{\eta} + B_{\pi} + B_{NN} =1$.
We consider the branching ratios $B_\eta$, $B_\pi$, and $B_{NN}$ as parameters for the simulation 
because of the uncertainty of the strength of each process
and assume two sets of values as $(B_\eta, B_\pi, B_{NN})=(0.2, 0.2, 0.6)$, and $(0.4, 0.4, 0.2)$ in this article.
The results shown in Fig.~\ref{fig:3}~(right) and Table~\ref{tab:1} correspond to the case with $(B_\eta, B_\pi, B_{NN})=(0.0, 0.0, 1.0)$.
We show in Fig.~\ref{fig:SBR} the simulated distribution of the emitted protons for the $\eta^\prime$ absorption as in Fig.~\ref{fig:3}~(right) for the two-body process only.
As in the case of the two-body absorption, the emitted protons  have an approximately uniform angular distribution in the laboratory frame in the one-body processes, too. 
As we can see in Fig.~\ref{fig:SBR}, by increasing the branching ratios of one-body process $B_\eta$ and $B_\pi$, the peak momentum of the emitted protons becomes smaller as expected.
The number of $\eta^\prime$ mesic nuclei, which produce the emitted proton(s) satisfying the cut conditions, are obtained by the simulation in the similar ways as described in Section~\ref{sect:3.1} for all decay channels.
This number is the same as that of corresponding forward deuterons in the signal process in the semi-exclusive measurements and used to evaluate the improvement factor $f_{S/B}$.

We show the results of $f_{S/B}$ in Tables~\ref{tab:2} and \ref{tab:3} for two sets of branching ratios.
As we can see from these tables, the $f_{S/B}$ factors for the strong cut conditions for $p_p$ decrease rapidly
as $B_\eta$ and $B_\pi$ increase because the number of protons with high momentum $\sim 1$~GeV/$c$ from two-body $\eta^\prime$ absorption decreases.
For example, the $f_{S/B}$ factor with the cut condition $p_p > 0.8$~GeV/$c$ and $\theta_p > 90^\circ$ is $2.1 \times 10^2$ for $B_{NN} = 1.0$ as shown in Table~\ref{tab:1}, which becomes $f_{S/B} = 1.3 \times 10^2$ for $B_{NN} = 0.6$ and $f_{S/B} = 5.6 \times 10^1$ for $B_{NN} = 0.2$.
The $f_{S/B}$ factor tends to decrease more sharply for stronger cut conditions for $p_p$ as $B_{NN}$ decreases.
In addition, we notice that the $f_{S/B}$ factors also become smaller gradually for larger $B_\eta$
and $B_\pi$ even for the weaker cut conditions for $p_p$.
This is because the increase in the protons with lower momentum for larger $B_{\eta}$ and $B_{\pi}$ results in the increase in the signal protons excluded under weak cut conditions.
In any cases, regardless of the value of the branching ratios, we find that we can expect the large $f_{S/B}$ factors $(\gg 1)$ for the semi-exclusive measurements with protons with large momentum at backward angles.

\begin{table}[!htb]
\caption{\label{tab:2} 
Same as Table~\ref{tab:1} except for inclusion of $\eta^\prime N \to \eta N$ and $\eta^\prime N \to \pi N$ decay processes in addition to $\eta^\prime NN \to NN$ with assumed branching ratios 
$(B_{\eta}, B_{\pi}, B_{NN}) = (0.2, 0.2, 0.6)$. 
The results with the cut conditions $0.3 < p_p $ and $0.5 < p_p$ are 
added in this table to focus on the lower momentum region.
}
\begin{tabular}{l||ccccccc}
\hline
 &   \multicolumn{5}{c}{Proton momentum cut  [GeV/$c$]} \\ 
Proton angle cut &~~~$0.3<$~~~ & ~~~$0.5<$~~~ &  ~~~$0.6<$~~~ &   ~~~$0.7<$~~~ &  ~~~$0.8<$~~~ & ~~~$0.9<$~~~ & ~~~$1.0<p_p$~~~  \\
\hline
$\ ~0.50>  \cos \theta_p$ & $2.4\times 10^0$  &   $3.2\times 10^0$ &  $1.2\times 10^1$ & $2.8\times 10^1$ & $6.9\times 10^1$   & $1.7 \times 10^2$ & $4.2 \times 10^2$ \\
 $\ ~0.25>  $ & $2.7\times 10^0$   & $3.7\times 10^0$ & $1.6 \times 10^1$ & $3.9\times 10^1$ &   $1.0 \times 10^2$ & $2.6 \times 10^2$ &  $7.0 \times 10^2$ \\
$\ ~0.00>  $ & $3.1\times 10^0$  & $4.2\times 10^0$ & $2.0 \times 10^1$ & $5.0\times 10^1$ &  $1.3 \times 10^2$ &  $3.5 \times 10^2$ & $1.1 \times 10^3$ \\
$-0.25>$ & $3.3\times 10^0$ &  $4.4\times 10^0$ & $2.2 \times 10^1$ & $5.8\times 10^1$ &  $1.6 \times 10^2$  & $4.2 \times 10^2$ & $1.3 \times 10^3$ \\
$-0.50>$ & $3.4\times 10^0$ &   $4.5\times 10^0$ & $2.4 \times 10^1$  & $6.3\times 10^1$ &  $1.7 \times 10^2$  & $4.9 \times 10^2$ & $1.6 \times 10^3$ \\
$-0.75>$ & $3.3\times 10^0$ &   $4.4\times 10^0$ & $2.6 \times 10^1$ & $7.0\times 10^1$ &  $1.9 \times 10^2$ & $6.0 \times 10^2$ & $2.5 \times 10^3$ \\
\hline
\hline \\
 & $0.3 < p_p < 0.5$& & $0.5 < p_p < 0.7$ & &$0.7 < p_p < 0.9$ & & $0.9< p_p < 1.1$    \\
\hline
$\ ~0.0>  \cos \theta_p$~~~~~ & $ 1.1 \times 10^{0}$ & & $3.0 \times 10^0$ & & $1.7 \times 10^1$ & &$ 2.6 \times 10^2$   \\
\hline
\end{tabular}
\vspace{3mm}
\caption{\label{tab:3}
Same as Table~\ref{tab:2} except for the assumed branching ratios $(B_{\eta}, B_{\pi}, B_{NN}) = (0.4, 0.4, 0.2)$. 
}
\begin{tabular}{l||ccccccc}
\hline
 &   \multicolumn{5}{c}{Proton momentum cut  [GeV/$c$]} \\ 
Proton angle cut &~~~$0.3<$~~~ & ~~~$0.5<$~~~ &  ~~~$0.6<$~~~ &  ~~~$0.7<$~~~ &  ~~~$0.8<$~~~ & ~~~$0.9<$~~~ & ~~~$1.0<p_p$~~~  \\
\hline
$\ ~0.50>  \cos \theta_p$ & $2.1\times 10^0$  &   $2.5\times 10^0$ &   $8.2 \times 10^0$ & $1.5\times 10^1$ & $2.9\times 10^1$   & $6.2 \times 10^1$ & $1.5 \times 10^2$ \\
 $\ ~0.25>  $ & $2.3\times 10^0$   & $2.8\times 10^0$ & $1.1 \times 10^1$ &  $2.1\times 10^1$ &   $4.3 \times 10^1$ & $9.4 \times 10^1$ &  $2.4 \times 10^2$ \\
$\ ~0.00>  $ & $2.5\times 10^0$  & $3.0\times 10^0$ &  $1.2 \times 10^1$ &  $2.6\times 10^1$ &  $5.6 \times 10^1$ &  $1.3 \times 10^2$ & $3.8 \times 10^2$ \\
$-0.25>$ & $2.7\times 10^0$ &   $3.2\times 10^0$ & $1.4 \times 10^1$ &  $2.9\times 10^1$ &  $6.6 \times 10^1$  & $1.5 \times 10^2$ & $4.5 \times 10^2$ \\
$-0.50>$ & $2.7\times 10^0$ &   $3.2\times 10^0$ & $1.5 \times 10^1$ & $3.2\times 10^1$ &  $7.2 \times 10^1$  & $1.8 \times 10^2$ & $5.6 \times 10^2$ \\
$-0.75>$ & $2.7\times 10^0$ &   $3.1 \times 10^0$ & $1.6 \times 10^1$ &  $3.5\times 10^1$ &  $7.9 \times 10^1$ & $2.2 \times 10^2$ & $8.7 \times 10^2$ \\
\hline
\hline\\
 & $0.3 < p_p < 0.5$& & $0.5 < p_p < 0.7$ & &$0.7 < p_p < 0.9$ & & $0.9< p_p < 1.1$   \\
\hline
$\ ~0.0>  \cos \theta_p$~~~~~ & $1.2 \times 10^{0}$ & &$3.7 \times 10^0$ & & $1.4 \times 10^1$ & & $ 9.6 \times 10^1$   \\
\hline
\end{tabular}
\end{table}

\section{Conclusion}\label{sec:4}
We have studied theoretically the semi-exclusive $^{12}$C($p,dp$)$X$ reaction for the formation of $\eta^\prime$ mesic nuclei. In the last experiments of the $^{12}$C($p,d$)$X$ reaction~\cite{n-PRiMESuper-FRS:2016vbn,e-PRiMESuper-FRS:2017bzq,Sekiya:2025hwz}, we found that it was extremely difficult to identify the peak structure in the excitation energy spectra because of the huge background. In order to reduce the background and to observe possible peak structures, we considered in this article the semi-exclusive $^{12}$C($p,dp$)$X$ reaction where emitted protons from the $\eta^\prime$ absorption processes in the final state are detected in coincidence with forward going deuterons to select the events of the signal processes.

We have performed the numerical simulation by the microscopic transport model JAM and quantitatively evaluated the improvements of the $S/B$ ratio by the semi-exclusive measurements. We find that the semi-exclusive measurements of the final state protons at backward angles with high momenta can substantially improve the $S/B$ ratio, providing a significant improvement compared to the inclusive reaction for the observation of $\eta^\prime$ mesic states. 
Considering protons with momenta $p_p > 0.8$~GeV/$c$ and at angles $\theta_p > 90^\circ$ in the semi-exclusive measurement, the JAM simulation shows that the background is reduced significantly to 0.2\% relative to the inclusive reaction.
We demonstrate that the semi-exclusive measurements in coincidence with the particles emitted in $\eta^\prime$ absorption processes
are important in general for the $\eta^\prime$ mesic nucleus formation reactions as studied in this article.
The best experimental cut condition for the emitted protons can be determined by optimizing the various conditions including the statistics which is also affected by the cut conditions.
Applying the semi-excluive measurements in the actual experiment~\cite{Sekiya:2025hwz} a signal to background ratio of the order of 10\% has been achieved.

\section*{Acknowledgment}
We appreciate the fruitful discussions with Y. Nara. 
The work was partly supported by JSPS KAKENHI Grant Numbers 
JP18H01242, JP19K14709, JP20KK0070, JP23K03417, JP24H00238, JP24K07020, and JP25707018.

\appendix
\section{Relative Strength of Energetic Proton and/or Neutron Production in each $\eta^\prime$ absorption process}\label{sec:A}
In this appendix, we estimate the relative probabilities of the energetic proton(s) and/or neutron(s) production from each absorption process of the bound $\eta^\prime$ by nucleon(s) inside the nucleus. The produced energetic nucleons propagate inside the nucleus to the outside. This propagation process is investigated by the simulation separately.
The production and propagation of $\pi$ and $\eta$ are also estimated and simulated for one-body absorption.
All hadrons produced by the $\eta^\prime$ absorption processes can be the source of the proton emission outside the nucleus.

\subsection{One-body absorption}\label{appen.1b}
\begin{figure}[hbt!]
 \begin{center}
 \includegraphics[scale=0.6]{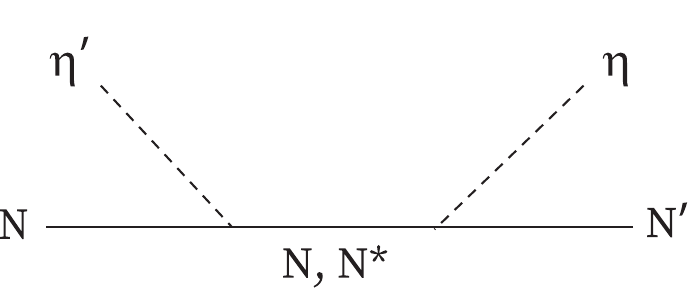}
\hspace{5mm}
 \includegraphics[scale=0.6]{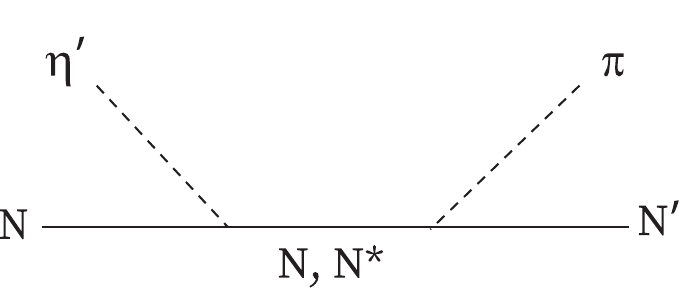}
  \caption{ The simple Feynman diagrams assumed in this article for (a) $\eta^\prime N \to \eta N^\prime$ and (b) $\eta^\prime N \to \pi N^\prime$ processes. $N$ and $N^\prime$ indicate proton and/or neutron, and $N^*$ the baryon resonances with isospin $1/2$. Because of isospin conservation, $\Delta$ resonances do not appear as the intermediate state in these diagrams.
 }
  \label{fig:one}
 \end{center}
\end{figure}

As one-body $\eta^\prime$ absorption processes, we consider here $\eta^\prime N \to \eta N^\prime$ and $\eta^\prime N \to \pi N^\prime$, where $N$ and $N^\prime$ indicate proton and/or neutron. We assume the simple diagrams as shown in Fig.~\ref{fig:one} to estimate the relative probabilities of the energetic proton and neutron production in each process.

First, we consider the $\eta$ production process shown in Fig.~\ref{fig:one}~(a). By considering the isoscalar nature of $\eta$ and $\eta^\prime$, we can expect the following relations between the transition amplitudes for $\eta$ production as,
\begin{equation}
T_{\eta^\prime p \to \eta p}  = T_{\eta^\prime n \to \eta n},  
\end{equation}
and
\begin{equation}
T_{\eta^\prime p \to \eta n}  = T_{\eta^\prime n \to \eta p} = 0 ,
\end{equation}
where the amplitude of the process $\eta^\prime N \to \eta N^\prime$ is written as $T_{\eta^\prime N \to \eta N^\prime}$.
Thus, within the $\eta^\prime N \to \eta N^\prime$ process, the probability $P_{\eta p}$ of the energetic proton production with $\eta$ can be estimated for the nucleus consisting of $Z$ protons and $N$ neutrons
as follows,
\begin{equation}
P_{\eta p} = \frac{Z |T_{\eta^\prime p \to \eta p}|^2}{ Z |T_{\eta^\prime p \to \eta p}|^2 + N |T_{\eta^\prime n \to \eta n}|^2 } = \frac{Z}{Z+N},
\label{eq:Petap}
\end{equation}
and the probability $P_{\eta n}$ of the energetic neutron production for $\eta$ production process is 
\begin{equation}
 P_{\eta n} = 1 - P_{\eta p} = \frac{N}{Z+N}.
\label{eq:Petan}
\end{equation}

Then, we consider the $\pi$ production process shown in Fig.~\ref{fig:one}~(b). Because of the isospin, we can expect the relation between neutral pion production amplitudes as, 
\begin{equation} 
|T_{\eta^\prime p \to \pi^0 p} | = |T_{\eta^\prime n \to \pi^0 n}|,
\end{equation}
and charged pion production amplitudes as,
\begin{equation} 
|T_{\eta^\prime n \to \pi^- p} | = |T_{\eta^\prime p \to \pi^+ n}|.
\end{equation}
In the simple diagrams shown in Fig.~\ref{fig:one}~(b), by considering the isospin factor of the charged and neutral pion vertices, we have
\begin{equation}
 |T_{\eta^\prime n \to \pi^- p}| = \sqrt{2} \ |T_{\eta^\prime n \to \pi^0 n}|,
\end{equation}
and
\begin{equation}
 |T_{\eta^\prime p \to \pi^+ n}| = \sqrt{2} \ |T_{\eta^\prime p \to \pi^0 p}|.
\end{equation}
Thus, 
within the $\eta^\prime N \to \pi N^\prime$ process, the probabilities $P_{\pi^{-} p}$ and $P_{\pi^{0} p}$ of the energetic proton production with the charged and neutral pion can be estimated for the nucleus consisting of $Z$ protons and $N$ neutrons
as follows,
\begin{align}
 P_{\pi^{-} p} &= \frac{ N |T_{\eta^\prime n \to \pi^{-} p}|^2 } { Z(|T_{\eta^\prime p \to \pi^{0} p}|^2 + |T_{\eta^\prime p \to \pi^{+} n}|^2 ) 
+ N (|T_{\eta^\prime n \to \pi^{0} n}|^2 + |T_{\eta^\prime n \to \pi^{-} p}|^2 )} \nonumber\\
&= \frac{ 2N} {Z(1+2) + N(1+2)} = \frac{2N} {3(Z+N)},
\label{eq:Ppimp}
\end{align}
and
\begin{equation}
 P_{\pi^{0} p}  = \frac{ Z |T_{\eta^\prime p \to \pi^{0} p}|^2 } { Z(|T_{\eta^\prime p \to \pi^{0} p}|^2 + |T_{\eta^\prime p \to \pi^{+} n}|^2 ) 
+ N (|T_{\eta^\prime n \to \pi^{0} n}|^2 + |T_{\eta^\prime n \to \pi^{-} p}|^2 )} 
= \frac{Z} {3(Z+N)},
\label{eq:Ppi0p}
\end{equation}
and the probabilities of the energetic neutron production for the charged and neutral $\pi$ production processes are
\begin{equation}
 P_{\pi^{+} n}  = \frac{ Z |T_{\eta^\prime p \to \pi^{+} n}|^2 } { Z(|T_{\eta^\prime p \to \pi^{0} p}|^2 + |T_{\eta^\prime p \to \pi^{+} n}|^2 ) 
+ N (|T_{\eta^\prime n \to \pi^{0} n}|^2 + |T_{\eta^\prime n \to \pi^{-} p}|^2 )} 
= \frac{2Z} {3(Z+N)},
\label{eq:Ppipn}
\end{equation}
and
\begin{equation}
 P_{\pi^{0} n}  = \frac{ N |T_{\eta^\prime n \to \pi^{0} n}|^2 } { Z(|T_{\eta^\prime p \to \pi^{0} p}|^2 + |T_{\eta^\prime p \to \pi^{+} n}|^2 ) 
+ N (|T_{\eta^\prime n \to \pi^{0} n}|^2 + |T_{\eta^\prime n \to \pi^{-} p}|^2 )} 
= \frac{N} {3(Z+N)}.
\label{eq:Ppi0n}
\end{equation}

\begin{figure}[hbt!]
 \begin{center}
  \includegraphics[scale=0.6]{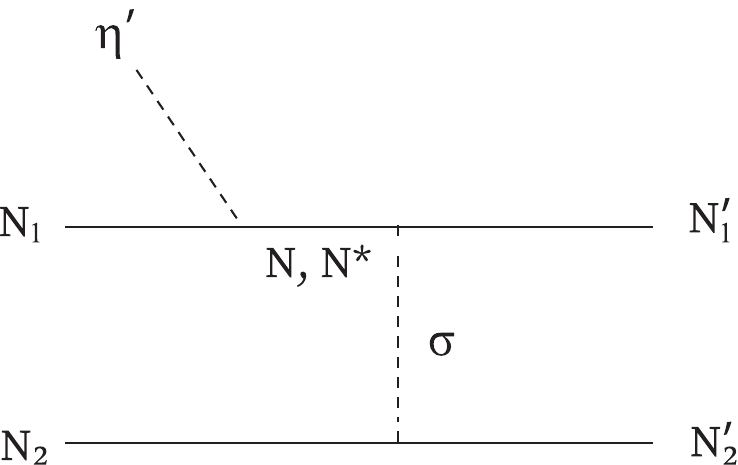}
  \caption{ The simple one-sigma exchange Feynman diagram assumed in this article for $\eta^\prime N_1 N_2 \to N_1^\prime N_2^\prime$ process. All $N$s indicate proton and/or neutron and $N^*$ the baryon resonances with isospin $1/2$. Because of isospin conservation, $\Delta$ resonances do not appear as the intermediate state in this diagram.
 }
  \label{fig:two}
 \end{center}
\end{figure}

\subsection{Two-body absorption}\label{appen.2b}
We assume the simplest diagram of $\sigma$\footnote{$\sigma$ is $f_0(500)$, and was $f_0(600)$ and $f_0(400-1200)$~\cite{PDG}.} exchange for two-body absorption processes as shown in Fig.~\ref{fig:two}. Because of the isoscalar nature of $\eta^\prime$, the resonances with isospin $1/2$ only contribute to the baryon intermediate state of this diagram.
By expressing the process in Fig.~\ref{fig:two} as $\eta^\prime N_1 N_2 \to N_1^\prime N_2^\prime$ and its amplitude as $T_{\eta^\prime N_1 N_2 \to N^\prime_1 N^\prime_2}$,
we can expect the following relations of the transition amplitudes of these processes as,
\begin{equation} 
T_{\eta^\prime pp \to pp}  = T_{\eta^\prime pn \to pn}  = T_{\eta^\prime np \to np}  = T_{\eta^\prime nn \to nn},
\end{equation}
and 
\begin{equation}
 T_{\eta^\prime pn \to np}  = T_{\eta^\prime np \to pn} =0,
\end{equation}
because of the isospin nature of $\eta^\prime$ and $\sigma$.
Thus, the two nucleon absorption processes of $\eta^\prime$ considered here are $\eta^\prime pp \to pp$, $\eta^\prime pn \to pn$, $\eta^\prime np \to n p$, and $\eta^\prime nn \to nn$. 
The amplitudes $T_{\eta^\prime pn \to np}$ and  $T_{\eta^\prime np \to pn}$ are evaluated to be zero simply because we only consider sigma exchange between nucleons as the dominant contribution. 

Thus, 
within the two-body absorption processes
the relative probability $P_{pp}$ of the energetic two proton production 
by the $\eta^\prime pp \to pp$ process from the $\eta^\prime$ mesic nucleus can be estimated 
for the nucleus consisting of $Z$ protons and $N$ neutrons
as follows,
\begin{align}
  P_{pp} &= \frac{Z(Z-1) |T_{\eta^\prime pp \to pp}|^2 }{ Z(Z-1)|T_{\eta^\prime pp \to pp}|^2 + N(N-1)|T_{\eta^\prime nn \to nn}|^2 + NZ|T_{\eta^\prime np \to np}|^2 + ZN|T_{\eta^\prime pn \to pn}|^2} \nonumber\\
&= \frac{Z(Z-1)}{(Z+N)(Z+N-1)},
\label{eq:Ppp_2N}
\end{align}
where we neglect the possible interference effects between 
the $\eta^\prime np \to np$ and $\eta^\prime pn \to pn$ processes.
Similarly the probability $P_{nn}$ of the energetic two neutron production is estimated as follows,
\begin{equation}
 P_{nn} = \frac{N(N-1)}{(Z+N)(Z+N-1)},
\label{eq:Pnn_2N}
\end{equation}
and the probability $P_{pn}$ of the energetic one-proton and one-neutron production by the $\eta^\prime n p \to np$ and $\eta^\prime pn \to pn$ processes is expressed as,
\begin{equation}
 P_{pn} = \frac{NZ + ZN}{(Z+N)(Z+N-1)} = \frac{2 ZN}{(Z+N)(Z+N-1)}.
\label{eq:Ppn_2N}
\end{equation}
As a consequence, on average 1.1 proton is emitted in the decay of $\eta^\prime \otimes {}^{11}\text{C}$ via the two-body absorption process $\eta^\prime N_1 N_2 \to N^\prime_1 N^\prime_2$.

The formulae given in Eqs.~\eqref{eq:Petap}, \eqref{eq:Petan}, 
\eqref{eq:Ppimp}, \eqref{eq:Ppi0p}, \eqref{eq:Ppipn}, \eqref{eq:Ppi0n}, 
\eqref{eq:Ppp_2N}, \eqref{eq:Pnn_2N}, and \eqref{eq:Ppn_2N}
only give the relative intensities within a given decay mode and not the relative intensity of the respective decay modes. For the single meson decays the strength is proportional to the nuclear density while the intensity of the two-body absorption process scales with the square of the nuclear density.

\section{Summary of the variables appeared in the explanation of the simulation}\label{sec:B}
We show in this appendix the compilation of the variables appeared in the explanation of the simulation in this article. 

\begin{table}[!htb]
\begin{tabular}{ll}
\hline\hline
(I) ~~$\eta^\prime N \to \eta N$ & ~  \\ 
(II) ~$\eta^\prime N \to \pi N$ &  one-body and two-body processes of $\eta^\prime$ absorption  \\
(III) $\eta^\prime NN \to NN$  & ~ \\ \hline
$B_\eta$, $B_\pi$, $B_{NN}$ 
& Assumed decay branching ratios of the $\eta^\prime$ mesic nucleus by the $\eta^\prime$ absorption processes (I), (II), and (III).\\\hline
$P_{\eta p}$ & Probability of the production of energetic PROTON with $\eta$ in nucleus from the bound $\eta^\prime$ by the process (I)  \\
& estimated in Appendix~\ref{appen.1b}. \\ \hline
$P_{\eta n}$ & Same as $P_{\eta p}$ except for the production of energetic NEUTRON.  
\\\hline
$P_{\pi^{-} p}$ & Same as $P_{\eta p}$ except for $\pi^{-}$ by the process (II). \\\hline
$P_{\pi^{0} p}$, $P_{\pi^{+} n}$, $P_{\pi^{0} n}$ & Same as $P_{\pi^{-} p}$ except for different charge states. 
 \\\hline
$P_{pp}$ & Probability of the production of two energetic PROTONS in nucleus from the bound $\eta^\prime$ by the process (III) \\
 & estimated in Appendix~\ref{appen.2b}. \\\hline
$P_{nn}$ & Same as $P_{pp}$ except for the production of two energetic NEUTRONS. \\\hline
$P_{pn}$ & Same as $P_{pp}$ except for the production of one energetic PROTON and one energetic NEUTRON. \\ \hline
\hline
 \end{tabular}
\end{table}








\end{document}